\documentclass[12pt]{elsarticle}
\usepackage{lineno,hyperref}
\modulolinenumbers[5]
\usepackage[paperwidth=8.5 in, paperheight=11 in,twoside,textwidth=6.5 in,left=1 in, right = 1 in, top=1 in]{geometry} 

\makeatletter
\def\ps@pprintTitle{%
 \let\@oddhead\@empty
 \let\@evenhead\@empty
 \def\@oddfoot{\centerline{\it{Preprint, 8 December 2021}}}%
 \let\@evenfoot\@oddfoot}
\makeatother

\usepackage{subfig} 
\usepackage{epstopdf} 
\usepackage{amsmath,amssymb,amsthm,amsfonts} 
\usepackage{varioref}
\usepackage{doi}
\usepackage{booktabs}
\usepackage{caption}
\usepackage{multirow}
\usepackage{longtable}
\usepackage{pdflscape}
\usepackage{xcolor}
\usepackage{lscape}

\usepackage{setspace} 

\usepackage{verbatim}

\usepackage{adjustbox}

\usepackage{hyperref}

\theoremstyle{definition}

\theoremstyle{plain}

\theoremstyle{plain}

\theoremstyle{plain}

\theoremstyle{plain}

\begin{document}
 
\title{The Nature of Losses from Cyber-Related Events: \\ Risk Categories and Business Sectors}

  \author[add1]{Pavel V. Shevchenko}
  
  \author[add1]{Jiwook Jang}
  
  \author[add1]{Matteo Malavasi}
  
  \author[add2]{Gareth W. Peters\corref{cor1}}
  
  \author[add5]{Georgy Sofronov}
  \author[add1]{Stefan Tr\"uck}
  
  \cortext[cor1]{Correspondence address. Department of Actuarial Mathematics and Statistics, School of Mathematical and Computer Sciences, Heriot-Watt University, Edinburgh, Scotland, EH14 4AS, UK. E-mail: garethpeters78@gmail.com}
  
  \address[add1]{Department of Actuarial Studies \& Business Analytics, Macquarie Business School, Macquarie University, Sydney NSW 2109, Australia}
  \address[add2]{Department of Statistics and Applied Probability, College of Letters \& Science, University of California Santa Barbara, Santa Barbara, California 93106 USA}

  \address[add5]{School of Mathematical and Physical Sciences, Faculty of Science and Engineering, Macquarie University, Sydney NSW 2109, Australia}

\begin{abstract}
 \noindent
In this study we examine the nature of losses from cyber-related events across different risk categories and business sectors. Using a leading industry dataset of cyber events,
we evaluate the relationship between the frequency and severity of individual cyber-related events and the number of affected records. We find that the frequency of reported cyber‐related events has substantially increased between 2008 and 2016. Furthermore, the frequency and severity of losses depend on the business sector and type of cyber threat: the most significant cyber loss event categories, by number of events, were related to data breaches and the unauthorized disclosure of data, while cyber extortion, phishing, spoofing and other social engineering practices showed substantial growth rates. 
Interestingly, we do not find a distinct pattern between the frequency of events, the loss severity, and the number of affected records as often alluded to in the literature. We also analyse the severity distribution of cyber-related events across all risk categories and business sectors. This analysis reveals that cyber risks are heavy-tailed, i.e., cyber risk events have a higher probability to produce extreme losses than events whose severity follows an exponential distribution. Furthermore, we find that the frequency and severity of cyber-related losses exhibits a very dynamic and time-varying nature. 

\vspace{0.5cm}


\end{abstract}

\maketitle

\section{Introduction}
\onehalfspacing
 \noindent 
  According to a recent estimate provided in the Global Risk Report by the World Economic Forum \cite{worldeconomicforum2020}, losses from cyber-related risks are expected to increase by up to US\$ 6 trillion in 2021. Due to the digitalisation of business and economic activities via the Internet of Things (IoT), cloud computing, mobile, blockchain and other innovative technologies, cyber risk is inherent and extreme. In general, one may refer to cyber risk as any risk of financial loss, disruption to operations, or damage to the reputation of an organisation due to failure of its information technology (IT) systems, as defined by the Institute of Risk Management (IRM) in \cite{irm2014a,irm2014b}. Financial losses from malicious cyber activities result from IT security/data/digital assets recovery, liability with respect to identity theft and data breaches, reputation/brand damage, legal liability, cyber extortion, regulatory defence and penalties coverage and business interruption. In the financial sector, cyber risk is classified by the Basel Committee on Banking Supervision \cite{BCBS2006} as a category of operational risk, for instance affecting  information and technology assets that can have consequences for the confidentiality, availability, and integrity of information and information systems \cite{cebula2010}.

In this study, we seek to understand from an empirical perspective, the relationships and dynamics of cyber risk loss processes. In order to achieve this we undertake a study of both the behaviour of the occurrence of events, known as frequency analysis, as well as the magnitude of events, known as severity analysis. As such, the study undertaken provides a thorough analysis of the frequency and severity of cyber-related events across different risk categories and business sectors. This is important, as cyber risk reaches all aspects of corporate, industry and government sectors at both the institutional level, down to an individual level. Furthermore, requirements on the need to mitigate, report and respond to emerging cyber threats differs in the regulatory intensity and requirements across each sector of industry and society. Therefore, one would expect differing patterns in the dynamics of cyber risk loss processes when viewed from differing industry and risk category perspectives. One may also expect the emergence of different trends in various industries to occur as new attack vectors or attacker capabilities are developed. Therefore, given the increasing importance of cyber threats to businesses, government and individuals, a statistical analysis of the nature of such events will help to develop appropriate risk management strategies and support investment into optimal mitigation of the risks. Without adequately quantifying the risks from different cyber threat categories for business sectors, it is impossible to manage the identified risks to be within acceptable levels.    
The frequency of malicious cyber activities is rapidly increasing, with the scope and nature dependent on an organisation's industry, size and location. According to the Allianz Global Risk Barometer 2021 \cite{allianz2021}, cyber incident (including cybercrime, IT failure/outage, data breaches, fines and penalties) is currently a top-three global business risk. It is therefore critical that corporations and governments focus on IT and network security enhancement. Unless public and private sector organisations have effective cyber security plans and strategies in place, and tools to manage and mitigate losses from cyber risks, cyber events have the potential to affect their business significantly, possibly damaging hard-earned reputations irreparably \cite{Mcshane2021}.

Due to the impact of COVID-19, business and economic activities will be also accelerated in cyber space, which could significantly increase the frequency and impact of cyber events around the globe, with alarming consequences for public and private sector organisations \cite{Lallie2021}. Concerns about higher frequency and severity of cyber catastrophes demand a re-examination of characterising cyber risks. A significant challenge ahead is to quantify individual cyber-related events in order to better understand the current and emerging risk landscape in cyber space and to minimise potentially catastrophic losses from cyber activities.

The lack of historical data on losses from cyber risk is another challenge to model the frequency and severity of individual cyber-related events \cite{Biener2015,eling2019actual,Gordon2003,wef2020}. For example, in Australia, it only became mandatory for breached organisations to notify their data breaches details in February 2018 (see \cite{Bill2016}). Many countries around the world are in a similar situation, such that often only very limited data on losses from cyber-related events is available. This makes the design of adequate models for the quantification of cyber risks very difficult.  

McShane et al. \cite{Mcshane2021} provide a comprehensive review of the literature on managing cyber risks, focussing in particular on work that is related to risk identification, risk analysis, and risk treatment. For each of these steps, an appropriate quantification of potential losses from cyber-related events is paramount to consider and operationalise. The modeling and quantification of cyber losses is also a key component required to incorporate cyber risk into an overall enterprise risk management process and to facilitate the prioritisation of investment decisions aimed at reducing the impact of cyber attacks. 

This study provides a thorough analysis of the frequency and severity of individual cyber-related events for different cyber threats and business sectors. 
To do this, we use one of the most comprehensive databases on losses from cyber-related threats provided by Advisen\footnote{Advisen is a leading provider of data, media, and technology solutions for the commercial property and casualty insurance market. Advisen’s proprietary datasets include “\emph{Cyber Loss Data}”, “\emph{Casualty Dataset}”, “\emph{Private D\&O Loss Data}”, “\emph{Public D\&O Loss Data}” and “\emph{Loss Insight}”.},  hereafter referred to as Advisen Cyber Loss Data (\url{https://www.advisenltd.com/data/cyber-loss-data/}). Romanosky \cite{Romanosky2016} examined an early version of this dataset (i.e. over 12,000 cyber events from 2004-2015) that includes data breaches, security incidents, privacy violations, and phishing crimes with a regression analysis. In contrast, the dataset analyzed in this study comprises over 132,126 cyber events from 2008-2020, affecting 49,496 organizations, with more than 80\% of the organizations represented in the dataset residing in the United States. This is both a more current analysis and we argue more complete in coverage than previous studies. This is important given the dynamic and non-stationary nature of cyber loss processes, as we will demonstrate in the analysis and studies undertaken.

 Another main focus of this article is to demonstrate the main features of cyber risk loss processes, by looking at cyber event severity and frequency, using the Advisen Cyber Loss Data. There are various statistical models for frequency and severity used in operational risk practice suitable for cyber risk; see e.g. monographs devoted to such models \cite{cruz2015fundamental, shevchenko2011modelling}. Cohen et al. \cite{Cohen2019} also suggest that cyber losses and non cyber losses from operational risk share a similar fundamental risk profile. This would suggest that modeling techniques that have been originally developed for operational risk modeling may also be adequate for cyber-related threats. Whilst in this work we undertake a purely empirical analysis, the results we present add insight into the nature of the statistical models (such as considered in \cite{malavasi2021cyber}) that will be suitable to capture cyber risk loss processes adequately and the challenges faced in trying to achieve this in a non-stationary emerging dynamic threat domain.

We focus on an empirical analysis of the frequency and severity of losses from cyber events. We firstly evaluate the relationship between the frequency and severity of individual cyber-related events and the number of affected records. We find that the frequency and severity of the events are not independent of the business sector and type of cyber threat. Secondly, analysing the severity distribution of cyber-related events across all risk categories and business sectors provides us that cyber risk types are heavy tailed, i.e., cyber risk events have a higher probability to produce extreme losses than events whose severity follows an exponential distribution (for an approachable book length discussion on heavy tails see, for example, \cite{resnick2007}).  We also summarise detailed findings from the analysis of Advisen Cyber Loss Data, which are beneficial for a better understanding of the nature of cyber risk.
We close this article with some key recommendations to  cyber risk management decision makers in private and public sector organizations.

\section{The Data}
\noindent 
In our empirical analysis of the structure of losses from cyber-related threats, we use the Advisen Cyber Loss Data, which is one of the most comprehensive datasets on cyber events. The dataset contains incidents collected from reliable and publicly verifiable sources, such as news media, governmental and regulatory sources, state data breach notification sites, and third-party vendors. For cyber loss data, Advisen adopts a granular classification based on the type of cyber risk threat.

\subsection{Classification of Cyber Risks}
\noindent Cyber risk involves a wide variety of risk factors and touches on nearly every sector of the public and private domains, presenting many facets, combining technical know-how with behavioral and cultural aspects \cite{jrc2018, peters2018understanding,jrc2019}. This multidimensional heterogeneity makes defining and classifying cyber-related events a non-unique task, to the point where a globally accepted and standardized classification of cyber risk is not yet achieved universally accross all industry domains and sectors of society. Instead there are classifications and taxonomies which have been developed from different industry perspectives \cite{BCBS2006,Biener2015,cebula2014,cebula2010,cro2014,cro2016,cyentia2016}. Furthermore, many public and private institutions have tried to produce classifications addressing the most prominent cyber risk aspects relevant to their own stakeholders \cite{nist2004,jrc2018,jrc2019,acsc2020}.

For our empirical analysis, we decided to follow the classification that has been suggested in the  Advised Cyber Loss Data.\footnote{Note that the proposed cyber loss event and business line categories are distinct from those defined by the Basel II regulatory framework \cite{BCBS2006} for operational risk. Thus, when working with a taxonomy distinct from that specified in regulation for the banking sector, financial institutions subject to such regulatory reporting requirements will have to consider carefully the mapping exercise to move from the Advisen taxonomy to the Basel II required reporting taxonomy. Basel II business lines are largely non-representative of the taxonomy adopted by Advisen which covers a much wider selection of sectors including the financial services. As such, the integration of other cyber loss data collections such as those collected over the last 15 years in the banking industry by consortiums such as, e.g., ORX https://managingrisktogether.orx.org/  should be carefully considered.}  This classification comprises the following 16 cyber risk categories.
\begin{itemize}
\setlength\itemsep{0em}
    \item \textbf{Privacy – Unauthorized Contact or Disclosure}: cases when personal information is used in an unauthorized manner to contact or publicize information regarding an individual or an organization without their explicit permission.
    
    \item \textbf{Privacy – Unauthorized Data Collection}: cases where information about the users of electronic services, such as social media, cellphones, websites, and similar is captured and stored without their knowledge or consent, or where prohibited information may have been collected with or without their consent.
    
    \item \textbf{Data – Physically Lost or Stolen}: situations where personal confidential information or digital assets have been stored on, or may have been stored on, computer, peripheral equipment, data storage, or printouts which has been lost,stolen, or improperly disposed of.
    
    \item \textbf{Data – Malicious Breach}: situations where personal confidential information or digital assets either have been or may have been exposed or stolen, by unauthorized internal or external actors whose intent appears to have been the acquisition of such information.

    \item \textbf{Data – Unintentional Disclosure}: situations where personal confidential information or digital assets have either been exposed, or may have been exposed, to unauthorized viewers due to an unintentional or inadvertent accident or error.
    
    \item \textbf{Identity – Fraudulent Use/Account Access}: identity theft or the fraudulent use of confidential personal information or account access in order to steal money, establish credit, or access account information, either through electronic or other means.
    
    \item \textbf{Industrial Controls and Operations}: losses involving disruption or attempted disruption to "connected" physical assets such as factories, automobiles, power plants, electrical grids, and similar (including “the internet of things”).
    
    \item \textbf{Network/Website Disruption}: unauthorized use of or access to a computer or network, or interference with the operation of same, including virus, worm, malware, digital denial of service (DDOS), system intrusions, and similar.

    \item \textbf{Phishing, Spoofing, Social Engineering}: attempts to get individuals to voluntarily provide information which could then be used illicitly, e.g. phishing or spoofing a legitimate website with a close replica to obtain account information, or sending fraudulent emails to initiate unauthorized activities (aka “spear phishing”).
    
    \item \textbf{Skimming, Physical Tampering}: use of physical devices to illegally capture electronic information such as bank account or credit card numbers for individual transactions, or installing software on such point-of-sale devices to accomplish the same goal.
    
    \item \textbf{IT – Configuration/Implementation Errors}: losses resulting from errors or mistakes which are made in maintaining, upgrading, replacing, or operating the hardware and software IT infrastructure of an organization, typically resulting in system, network, or web outages or disruptions.
    
    \item \textbf{IT – Processing Errors}: Losses resulting from internal errors in electronically processing orders, purchases, registrations, and similar, usually due to a security or authorization inadequacy, software bug, hardware malfunction, or user error.
    
    \item \textbf{Cyber Extortion}:  threats to lock access to devices or files, fraudulently transfer funds, destroy data, interfere with the operation of a system/network/site, or disclose confidential digital information such as identities of customers/employees, unless payments are made.
    
    \item \textbf{Denial of Service (DDOS)/System Disruption}: a disruption which is too widespread to be accounted to individual organizations/entities.
    
    \item \textbf{Digital Breach/Identity Theft}: widespread hacking or identity theft which targets a large number of companies, or individuals.
    
    \item \textbf{Undetermined/Other}.
    
\end{itemize}
The dataset is also classified using the following 20 business sectors proposed by the North American Industry Classification (NAIC) system \cite{naic2020}:
\begin{itemize}
\setlength\itemsep{0em}
\item Agriculture, Forestry, Fishing and Hunting 
\item Mining, Quarrying, and Oil and Gas Extraction 
\item Utilities 
\item Construction 
\item Manufacturing 
\item Wholesale Trade 
\item Retail Trade 
\item Transportation and Warehousing 
\item Information 
\item Finance and Insurance 
\item Real Estate and Rental and Leasing
\item Professional, Scientific, and Technical Services 
\item Management of Companies and Enterprises 
\item Administrative and Support and Waste Management and Remediation Services 
\item Educational Services 
\item Health Care and Social Assistance 
\item Arts, Entertainment, and Recreation 
\item Accommodation and Food Services 
\item Other Services (except Public Administration) 
\item Public Administration
\end{itemize}

The great conundrum for modelling and analysing cyber-related loss data is that whilst the prevalence of such events and their impact is seemingly growing over time, the access to national, standardised public domain data bases for such loss data is scarce and often prohibitively expensive to obtain \cite{wef2020,edwards2016,eling2019actual,eling2017}. Moreover, given that a widely accepted cyber risk definition and taxonomy does not exist universally across different sectors subject to different regulatory considerations and bodies, a dataset containing uniform and systematic information on cyber event severity and frequency is hard to find and to work with. As such, we believe that the dataset considered in this study represents an industry gold standard in this regard and therefore, should act as a meaningful representation of the current status of cyber risk loss process evolution.

\subsection{Preliminary Analysis}
\noindent The dataset analyzed in this study contains 132,126 cyber events from 2008-2020, affecting 49,496 organizations, with more than 80\% of the organizations represented in the dataset residing in the United States. It is important to also note that given the nature of cyber risk, the reporting requirements and the methods of data collection utilised in compiling this dataset under study, it is safe to assume that a relatively high share of events may not be recorded. It is well known in fact that enterprises and companies seldom and reluctantly report cyber-related events to avoid, among other things, a loss of reputation and trust from their counterparties. 

\begin{figure}[ht!]
    \centering
    \includegraphics[scale=0.4]{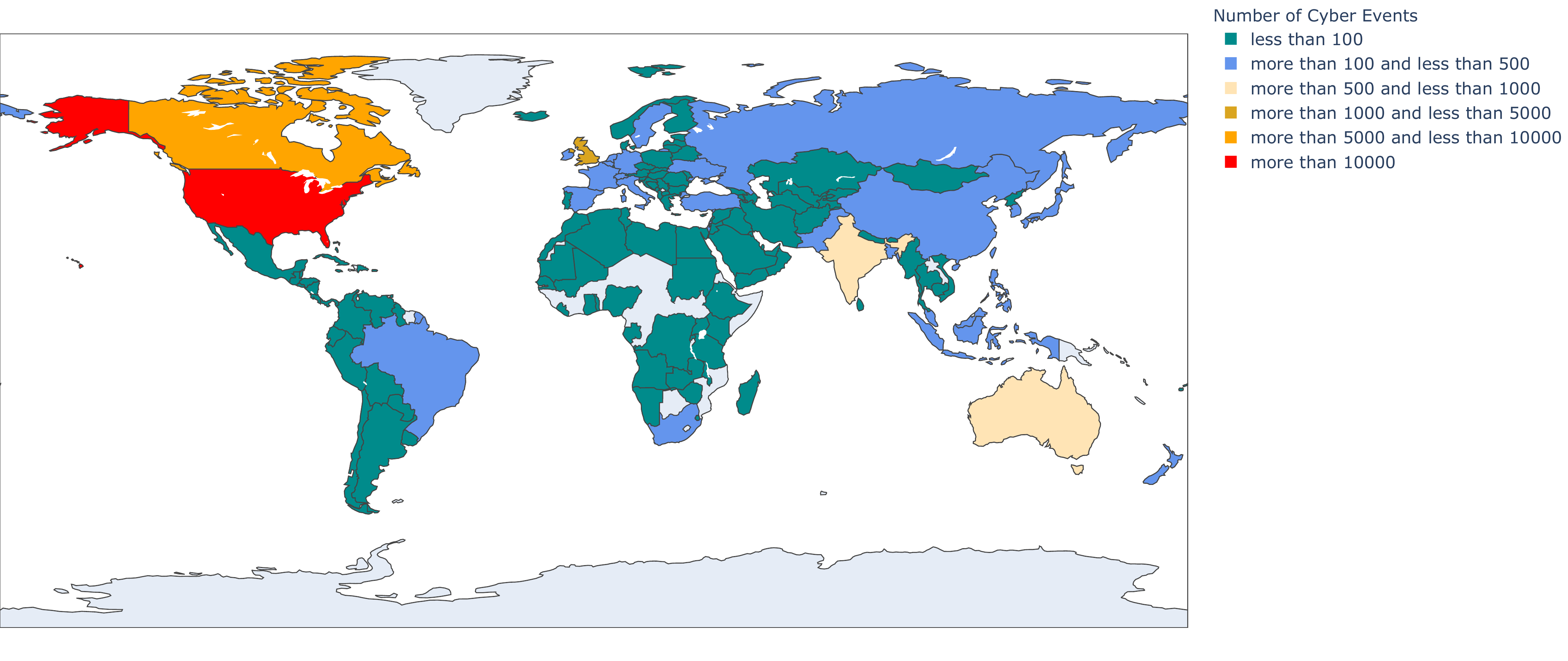}
    \caption{Number of cyber events by country during the period 2008-2020 across all loss categories .}
    \label{fig:2}
\end{figure}

Figure~\ref{fig:2} illustrates the number of events in the Advisen database by country. The vast majority of the recorded events in this database occurred in the United States (83.09\%), while only a minority of events is recorded for the entire European Union (2.65\%), Asia (3.17\%) or Oceania (1.04\%). As mentioned earlier, given the focus of the dataset on the United States, companies present in the dataset have been classified according to the NAIC system. In the following, we focus our analysis on non-zero losses in the dataset, i.e. 4,667 cyber events. Note that the share of these losses in the entire database is only 3.53\% of the total events. However, given our emphasis on the severity of cyber-related events, we have to rely on events where information on the magnitude of the loss was provided. 

Table~\ref{table:descriptive} provides descriptive statistics of non-zero losses for each cyber risk category. We find substantial differences for the number of non-zero loss events across the different risk categories. While we observe over 1,900 non-zero losses for the category \textit{Privacy - Unauthorized Contact or Disclosure} only six non-zero losses are observed for the category \textit{Industrial Controls} throughout the sample period. We also find heterogeneity in the magnitude of losses across the different categories. All risk categories exhibit a mean loss that is higher than the median, indicating that the loss distribution is skewed to the right, potentially exhibiting so-called heavy tails. In some cases this effect is so pronounced that the mean is more than 100 times higher than the median. Table~\ref{table:descriptive} also illustrates that losses from cyber-events typically have a very high standard deviation, positive skewness paired with high kurtosis. Overall, the descriptive statistics in Table~\ref{table:descriptive} also seem to confirm earlier results on cyber-related losses typically following so-called heavy-tailed distributions; see, e.g. \cite{maillart2010,edwards2016,eling2017}.

\begin{table}[h]
	\caption[Descriptive Statistics]{This table reports descriptive statistics of cyber risk related losses aggregated by categories. All dollar values are reported in million dollars. The losses exhibit great variability in terms of median and first four moments across the considered risk types. ``Digital Breach/Identity Theft", ``IT - Processing Errors", and ``Privacy - Unauthorized Data Collection" have the highest average loss amongst all cyber risk categories.}
	\label{table:descriptive}
	\hspace{-.5cm}\begin{tabular}{l|rrrrrr}
    	Risk Category                               & N & Mean &Median&StDev&Skew&Kurt \\
    	\hline
        Phishing \& Spoofing \&   Social Engineering& 202  &12.36 &0.57 &79.3   & 9.72 & 95.53 \\
        Privacy - Unauthorized Contact or Disclosure& 1916 &3.05  &0.03 & 23.8  & 31.75 & 1185.92 \\
        Data - Unintentional Disclosure             & 217  &1.34  &0.1  &8.81   & 12.73 & 172.27 \\
        Privacy - Unauthorized Data Collection      & 133  &46.77 &0.45 &434.07 & 11.18 & 124.5 \\
        Data - Malicious Breach                     & 858  &22.13 &0.5  &171.64 & 17.33 & 360.13 \\
        Identity - Fraudulent Use/Account Access    & 689  &1.2   &0.03 &6.55   & 10.28 & 124.79 \\
        Data - Physically Lost or Stolen            & 97   &23.91 &0.24 &202.14 & 9.63 & 91.16 \\
        Skimming \&  Physical Tampering             & 91   &1.72  &0.05 &6.08   & 6.14 & 42.59 \\
        IT - Processing Errors                      & 44   &76.55 &0.66 &264.77 & 5.32 & 29.25 \\
        IT - Configuration/Implementation Errors    & 63   &17.06 &0.8  &43.3   & 3.23 & 10.41 \\
        Network/Website Disruption                  & 181  &18.77 &0.16 &68.85  & 4.76 & 23.32 \\
        Cyber Extortion                             & 137  &0.52   &0.01 &2.78   & 6.86&48.24 \\
        Digital Breach/Identity Theft               & 11   &469.22 &30.0 & 1064.11 & 2.62& 5.24 \\
        Denial of Service (DDOS)/System Disruption  & 1    &0.39  &0.39 &-    & -& - \\
        Undetermined/Other                          & 21   &1.53  &0.65 &2.43   & 3.25& 10.68 \\
        Industrial Controls \& Operations           & 6    &30.7  &2.07 &62.39&1.78& 1.18 \\
	 	\hline	
		
	\end{tabular}
\end{table}

\section{Characterisation of Cyber Risks}

\subsection{Affected Records, Frequency and Severity of events}
\begin{figure}
    \hspace{-1.5cm}\includegraphics[scale=0.45]{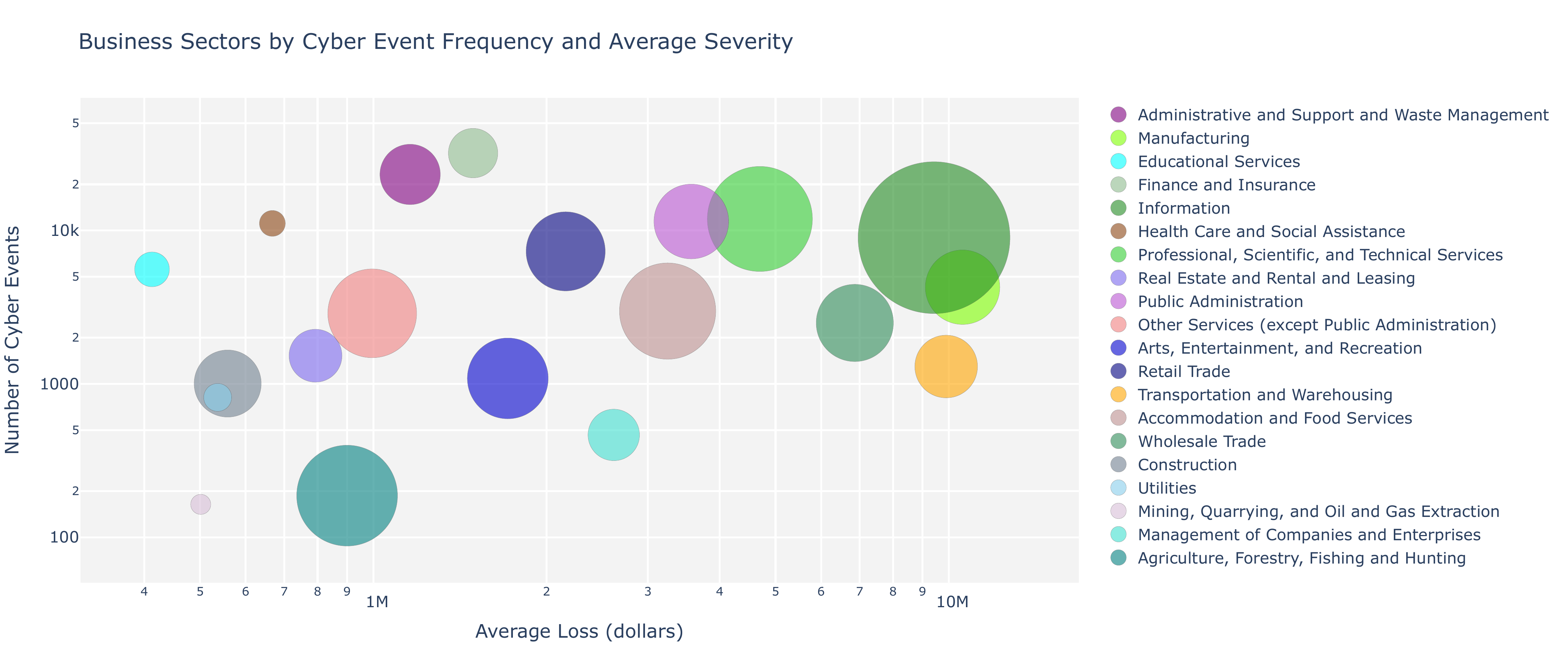}
    \caption{Frequency and severity of individual cyber-related events and the number of affected records (indicated by the size of the circle) across business sectors.}
    \label{fig:3}
\end{figure}

\noindent In a first step, we evaluate the relationship between the frequency and severity of individual cyber-related events and the number of affected records. Figure~\ref{fig:3} shows the business sector ranked by frequency and severity of cyber events. Each circle represents a business sector, and its area corresponds to the average number of records affected by a cyber event, i.e., the larger the circle the more records have been affected.

The sectors with the highest average cyber loss are: ``Information", ``Manufacturing", ``Transportation and Warehousing" and, ``Wholesale Trade". In terms of records affected: ``Information", ``Professional, Scientific, and Technical Services", ``Agriculture, Forestry, Fishing and Hunting", and ``Accommodation and Food Services". Figure~\ref{fig:3} also depicts the fact that monetary losses and the number of records affected vary across business sectors. Business sectors in the top right corner of the graph in Figure~\ref{fig:3} share some common features: they exhibit high average loss and high number of events, and a high average number of records affected (the bubbles have larger sizes than the sector in the top left corner of the graph). This seems to indicate that depending on the intrinsic nature of the business sectors, for some sectors there is a connection between a high number of records stolen which translates into high losses. However, for other sectors, a larger number of records does not necessarily translate into greater losses. For instance, records stolen in sectors such as ``Mining, Quarrying, Soil and Gas Extraction", ``Agriculture, Forestry, Fishing and Hunting", and ``Construction" have a lower monetary value than records stolen in ``Information" and ``Professional, Scientific and Technical Services".

\begin{figure}
    \hspace{-1.5cm}\includegraphics[scale=0.45]{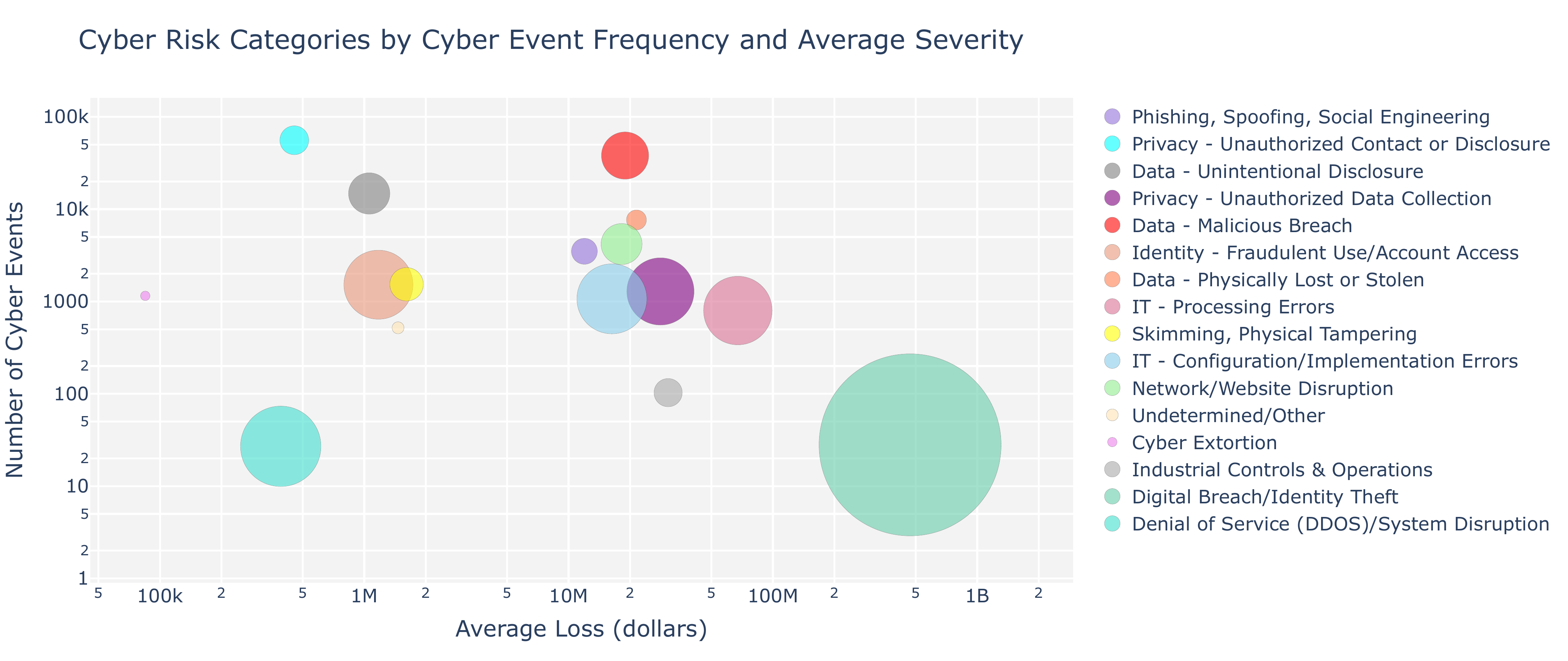}
    \caption{Frequency and severity of individual cyber-related events and the number of affected records (indicated by the size of the circle) across risk categories.}
    \label{fig:4}
\end{figure}

Figure~\ref{fig:4} shows the Advisen cyber risk threat types ranked by frequency and average severity. Each circle represents a business sector, and the area of the circle corresponds to the average number of records affected. The cyber risk type with the highest average loss and average number of records affected is  {``Digital Breach/Identity Theft"}. Looking at Figure~\ref{fig:4}, cyber risk types can be divided into three groups according to their average loss: 
\begin{enumerate}
    \item average loss lower than 2 million dollars: ``Cyber Extortion", ``Denial of Service(DDOS)/ System Disruption", ``Privacy - Unauthorized Contact or Disclosure", ``Data-Unintentional Disclosure", ``Identity Fraudulent Use/Account Access", and ``Skimming, Physical Tampering";
    \item average loss between 10 million dollars and 100 million dollars: ``Phishing, Spoofing, Social Engineering", ``IT-Configuration/Implementation Error", ``Network/Website Disruption", ``Data-Malicious Breach", ``Privacy-Unauthorized Data Collection" and ``IT-Processing Error";
    \item average loss greater than 100 million dollars: ``Digital Breach/Identity Theft".
\end{enumerate}

Overall, there seems to be no clear-cut relationship between the frequency of events, loss severity, and the number of affected records. The relationship depends also on the business sector and type of cyber threat.

\begin{figure}
    \hspace{-1.5cm}\includegraphics[scale=0.45]{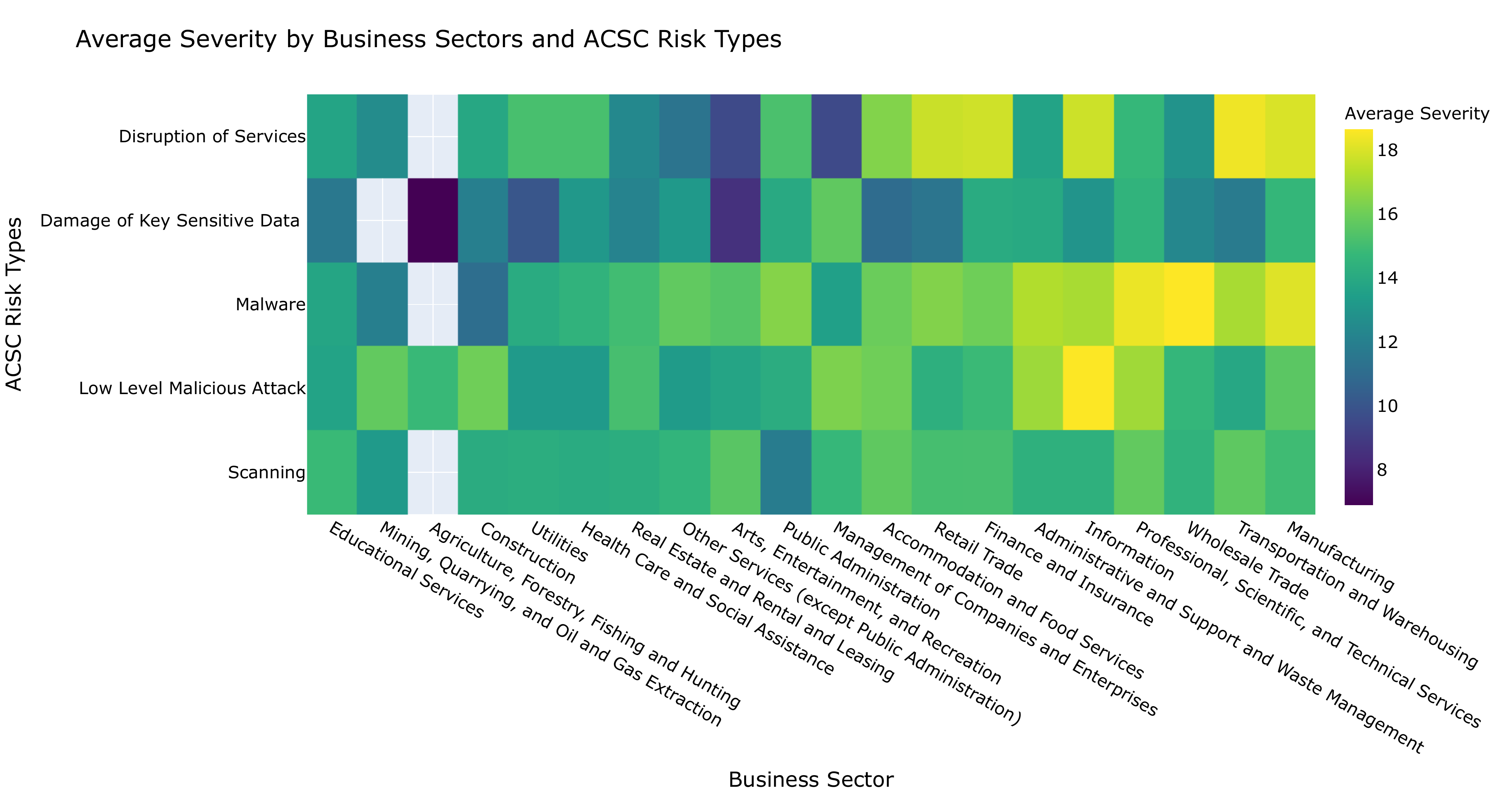}
    \caption{Average severity of cyber-related events by business sectors according to the ACSC classification. Sectors with high average losses show higher average severity in all the ACSC, than those with low overall average losses.}
    \label{fig:5}
\end{figure}

Figure~\ref{fig:5} proposes an adapted version of the categorisation matrix by the Australian Cyber Security Center (ACSC), in terms of business sectors and average cyber event severity \cite{acsc2020}. Figure ~\ref{fig:5} illustrates the heterogeneity in the severity of cyber events both for cyber risk type and business sector. In particular, sectors with high average loss in Figure~\ref{fig:3} report higher average losses for every ACSC cyber risk type, than those sectors with low average loss in Figure~\ref{fig:3}. This empirical fact emphasizes the dependence of monetary losses on specific company features, since companies operating in different business sectors have a different business model, a different internal structure and different levels of cyber risk resilience.

\subsection{The Frequency and Severity of Cyber Events}
\noindent In the following we examine the frequency and severity of cyber events in more detail.

\begin{figure}
    \hspace{-1.cm}\includegraphics[scale=0.45]{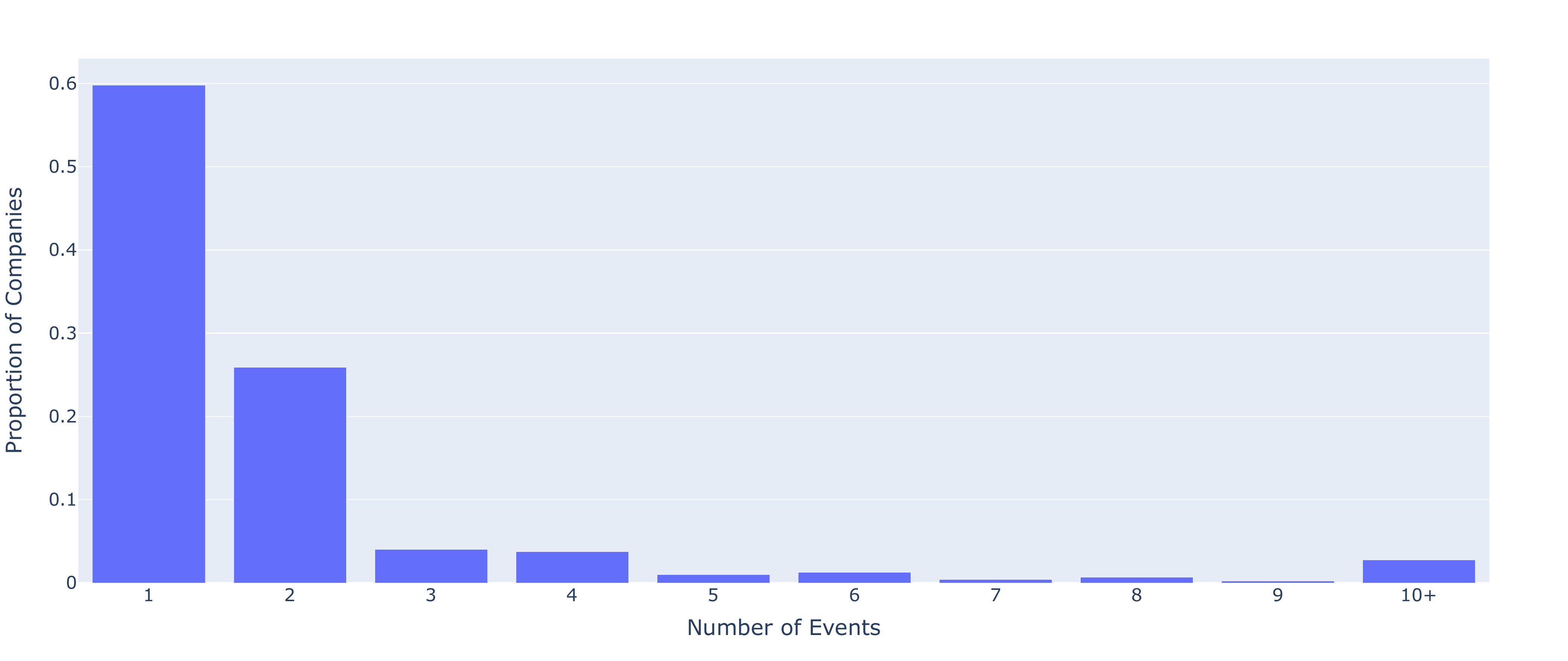}
    \caption{Distribution of the number of cyber-attacks per company.}
    \label{fig:6}
\end{figure}

Figure~\ref{fig:6} illustrates the distribution of the number of cyber attacks per company between 2008 and 2020. More than 40\% of the companies suffered from cyber crimes more than once during this period, with almost 3\% of the firms being affected more than 10 times. It is important to note that the dataset contains only information regarding cyber risk related events which have been publicly disclosed, and it would be safe to assume that the real number of events is much higher. 

\begin{figure}
    \hspace{-1cm}\includegraphics[scale=0.45]{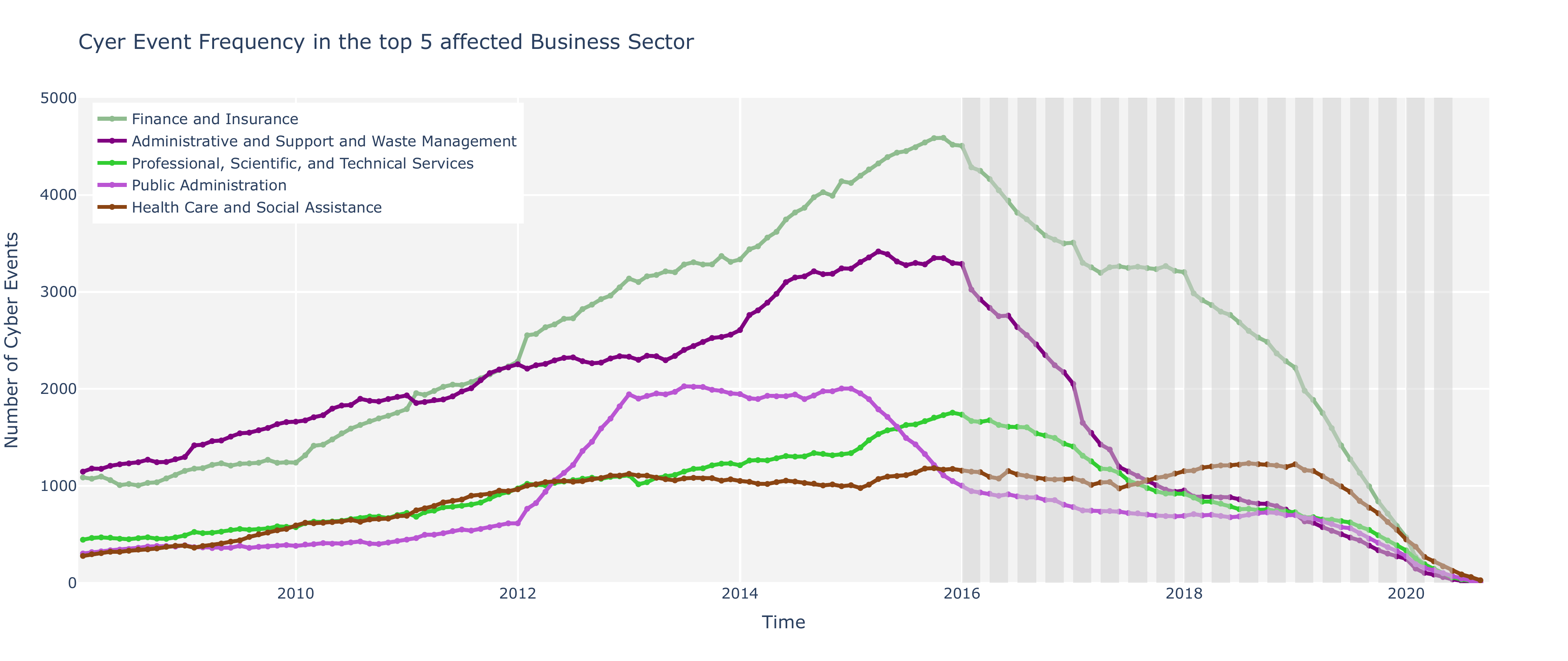}
    \caption{Number of cyber events for the five business sectors that were most affected.}
    \label{fig:7}
\end{figure}

Figure~\ref{fig:7} shows the number of cyber events for the five business sectors that were most affected. We find that the frequency of reported cyber-related events has substantially increased between 2008 and 2016 (4,800 reported events in 2008 versus 16,800 reported events in 2016). There is also a significant delay in the reporting of events that needs to be taken into account when drawing conclusions on the risks. As it can be seen from the graph, the number of events appears to be decreasing after 2016 (the period corresponding to the dashed area in the graph). Given that the decline is consistent across all business sectors, this seems to suggest the presence of a reporting delay, rather than a systematic improvement in cyber threat prevention, detection, and response mechanisms common for every business sector. Such reporting delay can be attributed to numerous factors, such as the reluctance of enterprises and companies to report cyber risk related events, and the data collection procedure employed by Advisen that abide to the United States of America Freedom of Information Act regulation \cite{US_FOIA}. As a matter of facts, while United States domiciled companies have a 60 day window between discovering the data breach and reporting it to affected parties, non-United States domiciled entities do not have such strict requirement. 
\begin{figure}
    \hspace{-1.5cm}\includegraphics[scale=0.45]{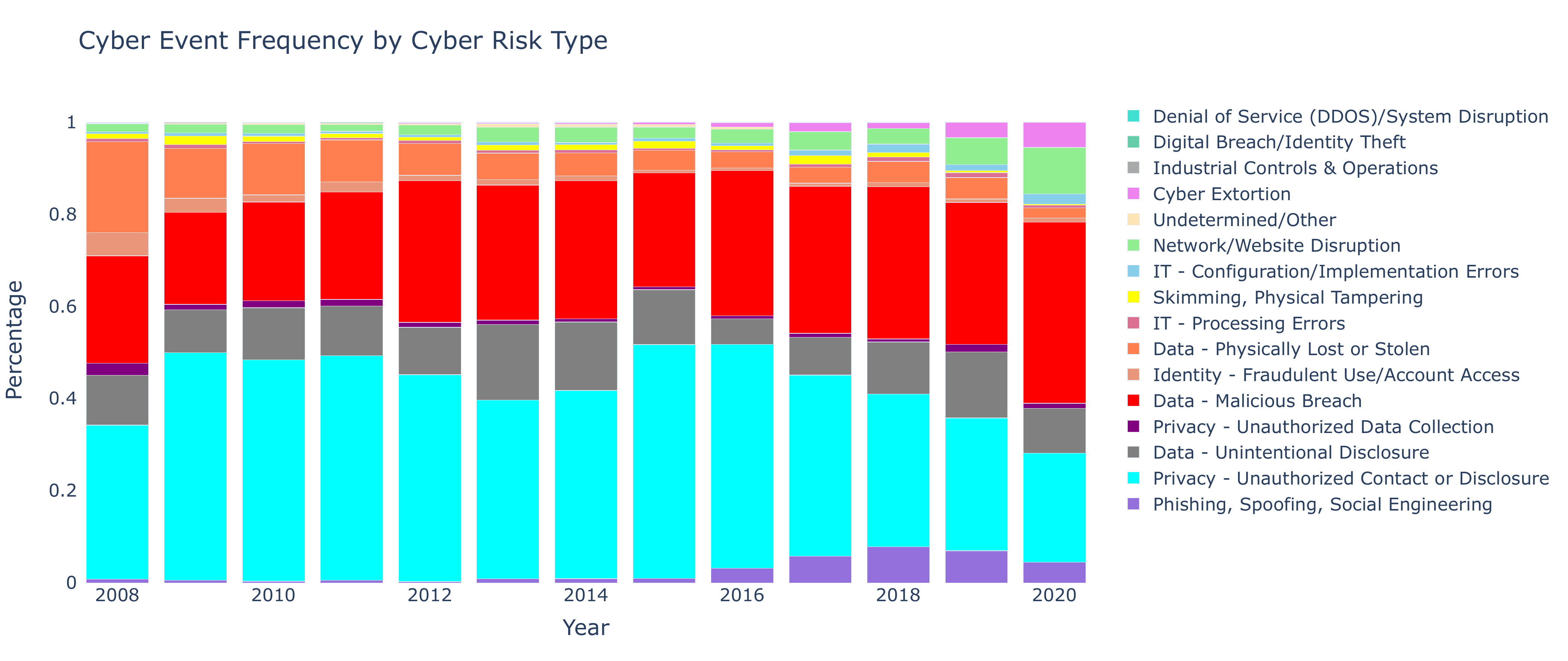}
    \caption{Share of cyber events for different cyber risk types for the period 2008 to 2020. Unintentional Disclosure, Data Malicious Breaches, Network/Website disruption, and Phishing, Spoofing and Social Engineering have become increasingly more common.}
    \label{fig:8}
\end{figure}

Figure~\ref{fig:8} illustrates the percentage of events that fall into a specific cyber risk category for the period 2008 to 2020. The figure also shows the dynamic nature of cyber risk, with substantially changing shares for different event types. In particular we find that cyber risk categories such as {``Data -- Unintentional Disclosure", ``Data -- Malicious Breaches", ``Network/Website Disruption"} have become increasingly more common since 2008. Moreover, in recent years, {``Cyber Extorsion"} and {``Phishing Spoofing and Social Engineering"} are on the rise, reflecting the capability of cyber criminals to adapt and create new forms of cyber threats. At the same time, the share of events for the category ``Data -- Physically Lost or Stolen" that played a major role in the years 2008-2011 has dropped significantly.

\begin{figure}
     \hspace{-1.5cm}
    \includegraphics[scale=0.45]{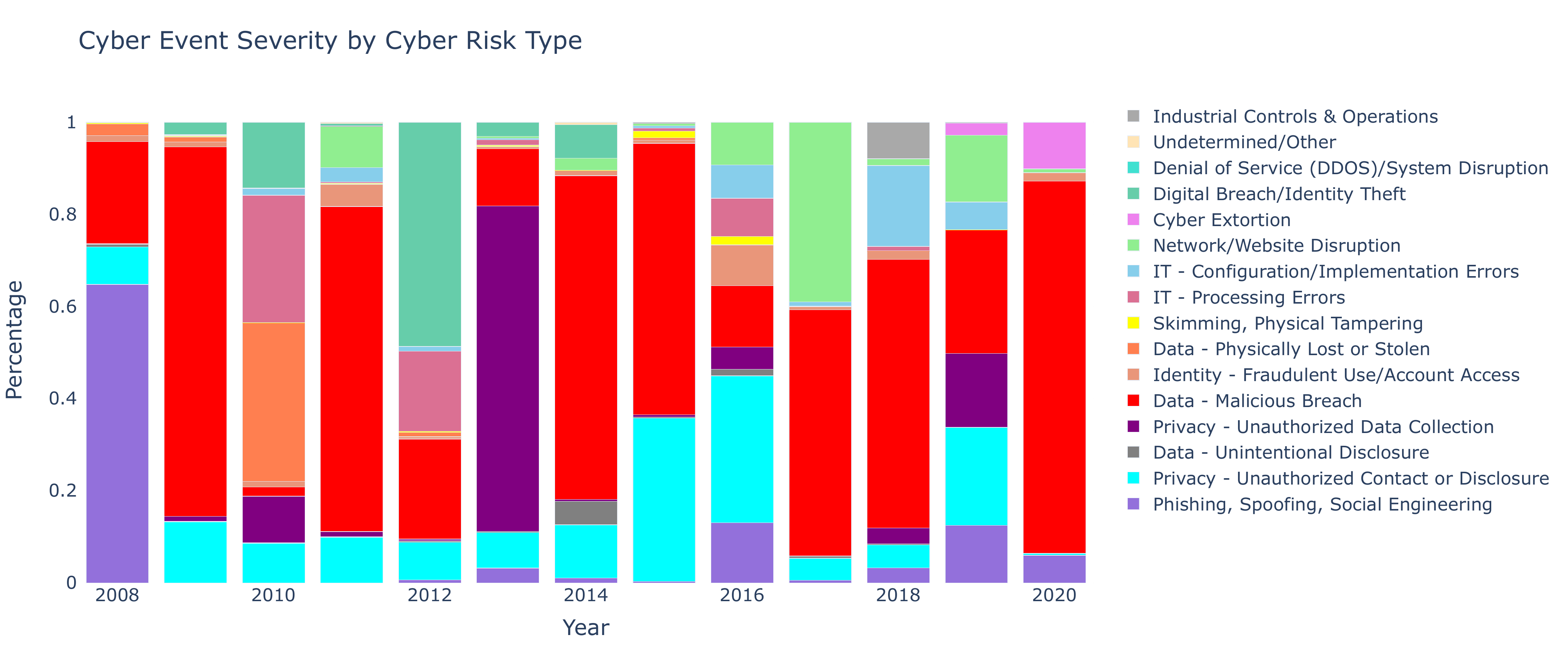}
    \caption{Share of total cyber-related losses that can be attributed to individual risk types for each year 2008-2020.}
    \label{fig:9}
\end{figure}

Cyber attacks are time varying in nature, and so are also the root causes of losses. Figure~\ref{fig:9} reports the share of total cyber-related losses that can be attributed to the different risk types for each year. Recall that for the frequency of different cyber risk categories we found a relatively clear structure as indicated by Figure~\ref{fig:8}. However for the severity of events, there is much more heterogeneity in the cyber risk categories across the time period. Recall that some losses from cyber events are extremely high, leading to a situation where in some years very few events (or even an individual event) can make up a relatively high percentage of the total loss during that year. Nonetheless, we find that losses from  {``Data-Malicious Breaches"} are typically among the highest, while this risk category can also be classified as the most severe risk type over the period 2017-2020. For other years, a high share of losses could be attributed to ``Phishing, Spoofing and Social Engineering" in 2008, ``Digital Breach/ Identity Theft" in 2012, ``Privacy – Unauthorized Data Collection" in 2013, and ``Network/Website Disruption" in 2017.
\begin{figure}
    \hspace{-1.5cm}
    \includegraphics[scale=0.45]{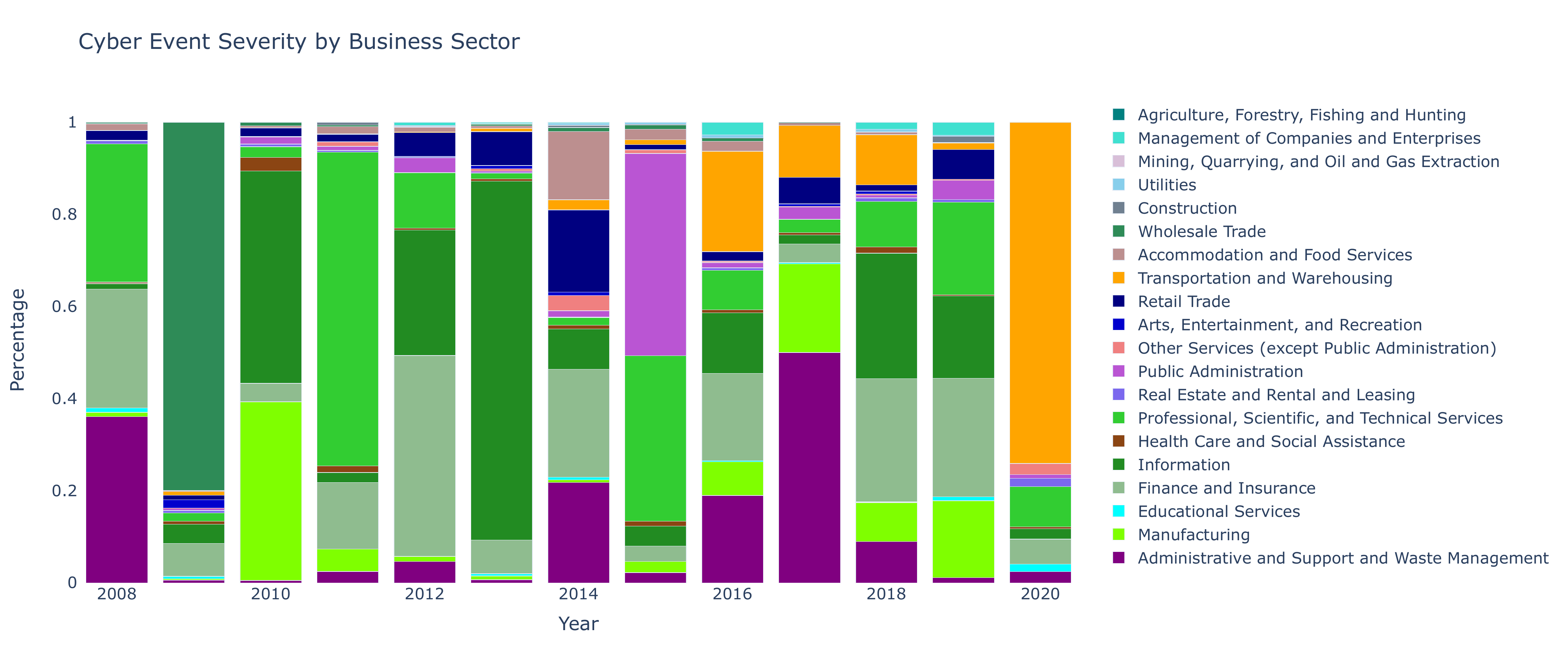}
    \caption{Share of total cyber-related losses that can be attributed to individual risk types for each year 2008-2020. Information, Professional Scientific and Technical Services, are Finance and Insurance are the most affected business sectors.}
    \label{fig:10}
\end{figure}

Not only does the nature of cyber risks change through time, but also companies in different business sectors suffer losses due to cyber events. Figure~\ref{fig:10} shows the share of total cyber-related losses that can be attributed to a specific business sector. Our results indicate that the {``Information"}, {``Professional Scientific and Technical Services"}, and the {``Finance and Insurance"} sectors typically seem to be among the most affected business sectors. However, this figure also illustrates that other sectors can be heavily affected by cyber events, for example ``Public Administration" in 2015 and ``Transportation and Warehousing" in 2020.

Overall, the frequency and severity of cyber-related losses exhibits a very dynamic and time-varying nature. While the occurrence of events seems to be dominated by certain risk categories, extreme losses occur in various cyber risk categories or business sectors. This behaviour also makes it particularly difficult to predict the nature or magnitude of losses from cyber-related events.

\subsection{The Severity Distribution}
\noindent Finally, we look at the severity distribution of cyber-related events across all risk categories and business sectors. Considering our sample, the majority of losses are typically relatively small, i.e. 85\% of events cause losses $<$\$2 million. However, we also observe a number of more extreme losses in the database: 5\% of losses exceed \$10 million, while 1.4\% of cyber-related losses exceed \$100 million, and 0.17\% of events cause losses that are $>$\$1 billion. Thus, the distribution of losses is clearly asymmetric, and contains some extreme observations.  

\begin{figure}
    \hspace{-1cm}\includegraphics[scale=0.45]{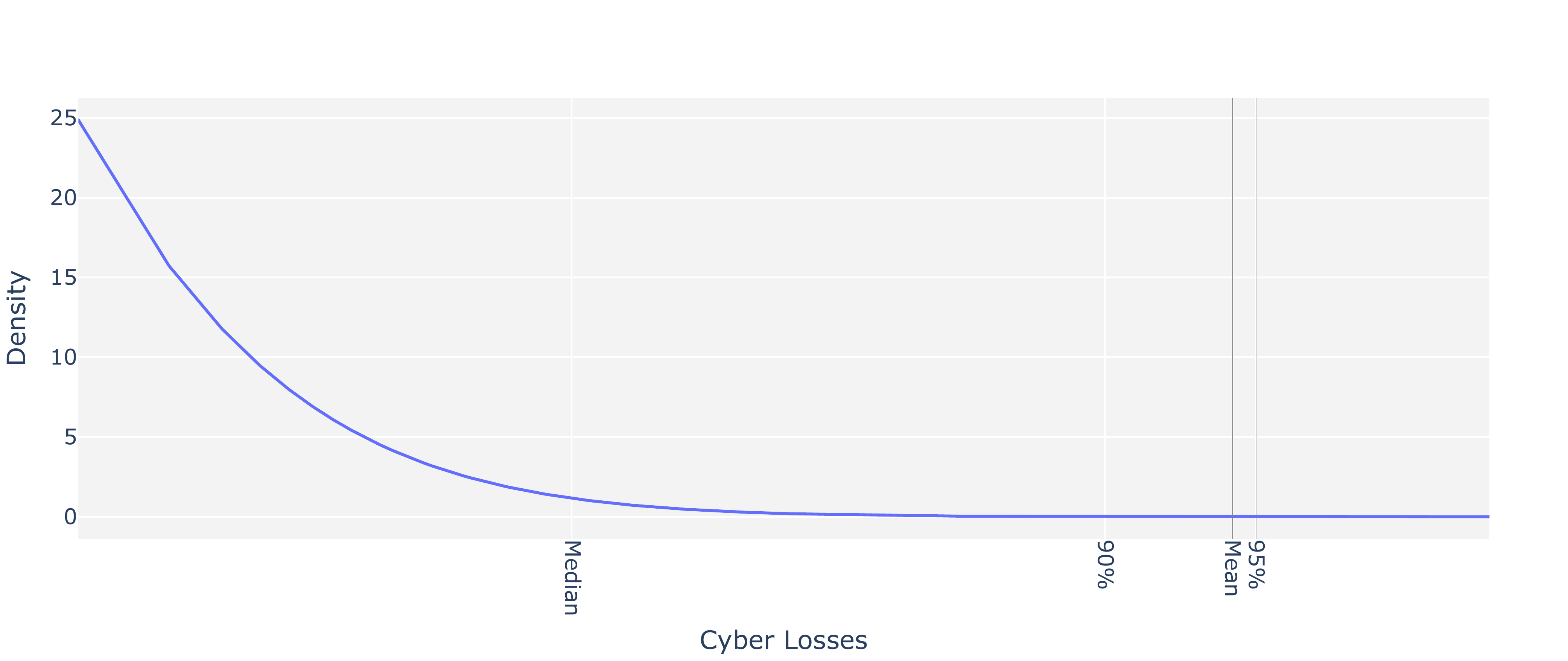}
    \caption{Probability density function of severity of cyber-related loss events, indicating the median of losses (0.108 million dollars), the 90\% quantile (6.364 million dollars), the mean (16.848 million dollars), and the 95\% quantile (20.180 million dollars).}
    \label{fig:11}
\end{figure}

Figure~\ref{fig:11} shows the fit of a lognormal probability density function to the severity of cyber events (blue line). The gray vertical lines correspond to, in ascending order of magnitude, the estimated median of losses from cyber events (0.108 million dollars), the estimated 90\% quantile (6.364 million dollars), the estimated mean (16.848 million dollars), and the estimated 95\% quantile (20.180 million dollars). Interestingly, the mean of the loss distribution is substantially higher (around 160 times) than the median. Note that the mean loss is even higher than the 90\% quantile of the distribution, confirming the substantial influence of a small number of very extreme events on the loss distribution. Moreover, the cumulative top 0.5\% highest losses yield approximately the same amount as the cumulative bottom 99.5\%. In Actuarial Science this phenomenon is called “heavy tails” and refers to the characteristic of certain probability distributions to allocate a small probability to very extreme events. In insurance, this is also reflected in the well known concept of “one loss causes ruin”, where the probability of one single event, having the potential to trigger losses so extreme that the company could fail to recover from is greater than zero. This is also consistent with results of statistical regression models considered in \cite{malavasi2021cyber}, where cyber event severity was found to follow a distribution so heavy tailed that, depending on risk types and company characteristics, it may present infinite mean.

\section{Conclusions and Policy Recommendations}
\noindent
In this article, we used a comprehensive dataset of cyber-related events to study the nature of cyber risk losses across different risk categories and business sectors. In particular we focused on the relationship between the frequency and severity of individual events and the number of affected records.

The studies undertaken demonstrated that over 60\% of companies that recorded cyber-related losses have suffered from cyber-attacks more than once in the period 2008-2020. This suggests that governance processes relating to mitigation of such events could be enhanced and that regulation and reporting around best practices as it emerges could help mitigate repeated events of the same nature from reoccurring.

It is also clear from the analysis that even with increasing scrutiny and increasing regulatory guidance that occurred in many industry sectors over the period of study, the rate of cyber crime has not abated. In fact, the frequency of reported cyber-related events has substantially increased between 2008 and 2016 (4,800 reported events in 2008, 16,800 reported events in 2016). Furthermore, the reporting of such events for modelling purposes could be improved as there appears to be a significant delay in the reporting of events that needs to be taken into account when drawing conclusions on the risks.

Furthermore, we found no distinct pattern or clear-cut relationship between the frequency of events, the loss severity, and the number of affected records. Contrary to assumptions often made in practice, the reported loss databases do not demonstrate a direct proportional relationship between total loss incurred from a cyber event and attributes from the event such as the number of compromised records (data records breached or stolen), the number of employees in a corporation or the number of units of a company affected. This finding shows that all companies, no matter the volume or size of data record can be susceptible to significant incurred loss from cyber events.

The frequency and severity of the events depend on the business sector and type of cyber threat. The most significant cyber loss event categories, by number of events, were ``Privacy -- Unauthorized Contact or Disclosure" and ``Data -- Malicious Breach". Data related breaches have become increasingly more common since 2008, while ``Cyber Extortion", ``Phishing, Spoofing and Social Engineering" practices also continue to increase, the pace at which malicious breach related events has occurred has now surpassed these other prominent categories of loss event risk type in recent years. In terms of business sectors,  the ``Information sector", ``Professional Scientific \& Technical Services", and ``Finance \& Insurance" have suffered most of the financial damage during the sample period 2008-2020.

Furthermore, the findings of the analysis re-affirmed that losses from cyber-related events are heavy-tailed. We found that the majority of losses were relatively small, i.e. 85\% of events cause losses less than \$2 million, while a very small number of events caused losses that were even greater than \$1 billion. Furthermore, the mean of the loss distribution (approximately \$16.8 million) was around 160 times higher than the median (approximately \$108,000). Thus, cyber losses are well represented by the expression “one loss causes ruin” adage. As such, in all categories of cyber loss type and in all sectors of the economy it was found that loss severity is often dominated by large individual events. 

Our findings also lead to a number of policy recommendations and suggestions for future work. Currently, data collection and databases on losses from cyber events have an unbalanced recording of samples with the strongest emphasis on the US data collection. However, cyber risk is international in nature affecting both industry as well as government agencies across all sectors of the economy. Similar to other risks such as credit or operational risk a concerted effort should be made to develop an adequate data collection and classification process for cyber-related risks in the international landscape. This should include detailed information on the event types, the failure modes that led to the loss events, and the components of the loss events broken down into categories. Thus, it would be beneficial for sector specific regulators to continue to develop a working taxonomy and reporting framework specific to the risk profiles and needs of different industry sectors. These should not be unified across all sectors of the economy but tailored to particular sectors to capture the heterogeneous nature of cyber risk data event types and loss behaviour profiles.

An effort should also be made in order to increase the awareness of cyber risk, even among small business and entities. Due to their heavy-tailed nature, losses from  cyber events can have catastrophic consequences against which even appropriate insurance policies might not be adequate to cover such losses. As cyber losses are currently significantly under-insured, there is a potential large exposure gap that could amount to tens of billions of dollars. This prblem should be addressed in the near future: with maturity of the reporting frameworks and consistency in the loss data collection, insurers will be able to more reliably price and design insurance contracts to mitigate some of the losses incurred from cyber risk, which in tandem with improving risk governance is a key component of risk transfer for this category of risk. Currently, the provided cyber insurance products have grown as a market exponentially, however their scope of coverage is extremely limited and bespoke in nature, making insurance premiums prohibitive to many markets. Consequently, at present, not all losses can be covered by insurance as there are extreme risks that pose a new challenge to the financial viability of insurers.

The dynamically changing nature of cyber risk dictates that enterprises, insurers and government agencies have to constantly update their cyber hygiene practices. Estimation of cyber risk based on historical losses is backward looking and thus there is a strong need to use scenario analysis to account for a constantly changing environment. There are various techniques in operational risk practice that allow to combine historical data and scenario analysis for estimation of risk frequency and severity;  see e.g. \cite{shevchenko2006structural,lambrigger2007quantification} or for a book length treatment \cite[chapters 14,15]{cruz2015fundamental}.

Main features of cyber risk we observe in the dataset and the fact that cyber risk types are heavy tailed call for further studies such as dependence analysis and tail asymptotic analysis. Such studies should be interesting for private and public sector organizations dealing with the ongoing challenges and new risk dynamics arising in the cyber space.

Overall, the conclusion of the findings from this analysis is that cyber loss processes are currently in a state of dynamical flux. This apparent non-stationarity in structural components of these processes can make it particularly challenging to make accurate predictions on the magnitude of losses from cyber‐related events. We conjecture that there is a combination of differing requirements by various industries in how and to what extent they must mitigate cyber losses. At the same time, there are varying requirements on the degree of reporting by industry. Therefore, as standardisation is introduced through regulation, this may consolidate the collection, reporting and understanding of cyber risk loss processes and with greater understanding and more universally applied mitigation strategies deployed, such loss processes may stabilise in their dynamics. This in turn allows for greater insight into how best to transfer risk associated with such loss processes through insurance and capitalisation.

\section*{Acknowledgements}
\noindent  This research has been conducted within the Optus Macquarie University Cyber Security Hub and funded by its Risk Management, Governance and Control Program.
\\

\section*{Conflict of interest statement} 
\noindent Authors declare no conflict of interest

\raggedright

\bibliography{bibliography}

\end{document}